# ExpertRank: A Multi-level Coarse-grained Expert-based Listwise Ranking Loss

Zhizhong Chen
Brown University
Providence, RI, USA

Carsten Eickhoff
Brown University
Providence, RI, USA

## ABSTRACT

The goal of information retrieval is to recommend a list of document candidates that are most relevant to a given query. Listwise learning trains neural retrieval models by comparing various candidates simultaneously on a large scale, offering much more competitive performance than pairwise and pointwise schemes. Existing listwise ranking losses treat the candidate document list as a whole unit without further inspection. Some candidates with moderate semantic prominence may be ignored by the noisy similarity signals or overshadowed by a few especially pronounced candidates. As a result, existing ranking losses fail to exploit the full potential of neural retrieval models. To address these concerns, we apply the classic pooling technique to conduct multi-level coarse graining and propose ExpertRank, a novel expert-based listwise ranking loss. The proposed scheme has three major advantages: (1) ExpertRank introduces the profound physics concept of coarse graining to information retrieval by selecting prominent candidates at various local levels based on model prediction and inter-document comparison. (2) ExpertRank applies the mixture of experts (MoE) technique to combine different experts effectively by extending the traditional ListNet. (3) Compared to other existing listwise learning approaches, ExpertRank produces much more reliable and competitive performance for various neural retrieval models with different complexities, from traditional models, such as KNRM, ConvKNRM, MatchPyramid, to sophisticated BERT/ALBERT-based retrieval models.

## 1 INTRODUCTION

Retrieval systems are the natural connection between various natural language processing techniques and large scale ranking applications. Natural language processing transforms the raw text information into useful mathematical representations which are understandable for the retrieval system. Neural retrieval models then extract all salient matching signals between the given query and document and produce a recommended list of the most relevant documents for the users [5, 10, 13, 23, 26, 34, 38].

How to rank document candidates more effectively is always at the core of information retrieval. In essence, the retrieval system attempts to sort the candidates based on the magnitude of semantic matching signals. However, the non-differentiability of the direct sorting breaks the back-propagation optimization of neural retrieval models. Hence, various learning to rank algorithms are proposed to aid model training. Pointwise ranking losses, such as cross entropy, treat relevance judgment as a simple classification problem [26]. In most realistic applications, relevant documents are much rarer than non-relevant candidates, which results in a difficult label imbalance problem. Also, it fails to incorporate inter-document information to generate optimization feedback. Recently, most neural retrieval models are trained in pairwise fashion, especially with margin ranking loss [14, 33]. RankNet [3] and LambdaRank [2] have been applied to many real-world ranking problems in the past successfully. Pairwise learning relies on the comparison between a pair of a relevant document and a non-relevant one. How to select these pairs is critical to the reliability of pairwise ranking losses. As a result, listwise learning approaches, such as ListNet [4], ListMLE [36], ApproxNDCG [27] and SoftRank [32], usually yield much more competitive performance. One major advantage of listwise learning is its innate ability to compare various candidates simultaneously on a large scale, which is much more consistent with the comparison-based ranking nature of information retrieval.

The mixture of experts (MoE) technique has been proposed for more than two decades [12, 15]. It successfully boosted many traditional machine learning methods, such as SVM [7] and Gaussian Processes [28]. The MoE approach constructs several "expert learners", each to focus on the different aspects of the problem. Then it employs a special gating network to control the importance of each expert. Ma *et al.* [21] apply it to model task relationships in multi-task learning. Zhao *et al.* [37] further build a multi-task ranking system for YouTube video recommendation. Its recent success on recommendation system encourages us to apply it for information retrieval. With the help of the MoE technique, we are able to combine various learning signals to further aid model training.

In this paper, we introduce a novel multi-level coarse-grained listwise ranking loss based on the mixture of experts (MoE) technique. Pooling layers with different pooling sizes are employed to compare various groups of documents at a local level to construct a series of filtered candidate lists. Coarse graining at different levels helps the candidates with moderately prominent semantic signals to stand out and provides more salient information to aid model learning. The traditional ListNet is then employed to build a "expert learner" for each coarse-grained candidate group. Finally, we apply the mixture of experts (MoE) technique to combine training information effectively across various experts. We refer to this mechanism as ExpertRank.

To demonstrate the generalizability and robustness of our method, we apply ExpertRank to five neural retrieval models: KNRM [5], ConvKNRM [38], MatchPyramid [23], and BERT/ALBERT-based retrieval models [8, 17, 26]. We conduct extensive experiments on the MS MARCO collection [24] and compare our performance with the existing listwise ranking losses. Furthermore, as one major drawback of listwise learning in general, the high computing memory requirement often prevents academic researchers from applying listwise learning to BERT/ALBERT-based retrieval models, which contain a notoriously huge number of parameters. We then propose a new two-step training process: efficient pointwise training followed by listwise re-training with fixed BERT/ALBERT embeddings. Finally, experiments on a news collection with limited query requests is used to further validate the effectiveness of ExpertRank in settings with limited training information.

The major novel contributions of this paper are (1) Building a new expert-based listwise ranking loss by applying the mixture of experts (MoE) technique in the domain of learning to rank. (2) Utilising the classic pooling technique to incorporate the profound physics concept of the multi-level coarse graining for information retrieval. (3) Empirical verification of robustness, generalizability and effectiveness of the proposed ranking mechanism with a general discussion on how to apply listwise learning to the powerful and sophisticated BERT/ALBERT-related retrieval systems.

The remainder of this paper is organized as follows. Section 2 reviews the related work on various learning to rank mechanisms and discusses the background of coarse graining in biophysics. Section 3 then establishes the formal network architecture of ExpertRank with detailed explanations of each major component. Section 4 further introduces some popular and effective neural retrieval models and explains how to apply ExpertRank to each system practically. Various experiments and corresponding performance comparison are presented and analyzed in Section 5. In Section 6, we study the effect of pooling size on ExpertRank performance. Finally, Section 7 concludes with a discussion of key findings and future directions of inquiry.

## 2 BACKGROUND AND RELATED WORK

Many neural retrieval models are constructed to extract salient semantic signals and to obtain a better understanding on how to quantify the relevance level between the given query and documents for user recommendation. Popular choices are KNRM [5], ConvKNRM [38], DeepMatch [20], MatchPyramid [23], DRMM [13], PACRR [10] and IRGAN [34]. Recently, due to its outstanding performance on several benchmark tasks, BERT and ALBERT models have drawn enormous attention for its solid applications in information retrieval [8, 17]. Various BERT/ALBERT-based retrieval models have been proposed to solve many realistic retrieval applications successfully [1, 26]. As model complexity keeps increasing, novel and better learning algorithms are strongly needed.

In general, learning to rank algorithms can be classified into three major categories: (1) Pointwise learning, which treats the ranking problem as a classic classification task. It has been applied to BERT-based retrieval models effectively due to its simplicity [26]. (2) Pairwise learning, which relies on a pair of relevant and non-relevant documents to compete with each other. Pairwise margin ranking loss [14, 33] is a popular choice for many retrieval models, such as KNRM [5], ConvKNRM [38] MatchPyramid [23] and DRMM [13]. RankNet [3] and LambdaRank [2] also gain wide popularity for many real-world ranking applications. (3) Listwise learning, which compares a list of document candidates simultaneously to find the relative ordering for the final prediction. Due to the comparison-based ranking nature of information retrieval, listwise learning usually provides much more significant and robust performance. Pointwise learning heavily depends on the model potential itself. For the complicated and powerful BERT/ALBERT-based models, pointwise learning can still yield reasonable performance. But its lack of inter-document comparison limits its ability to incorporate the relative ranking information for model training. On the other hand, pairwise learning only compares two candidates at the same time and deeply relies on the quality of document pairs. Various selective negative sampling methods have been proposed to aid candidate selection [6, 9, 22]. This may complicate the overall pipeline and limit the potential practical applications.

Listwise learning builds its analysis directly on the core of ranking problem, how to determine the relevance magnitude based on both query-document semantic connection and inter-document comparison. There are two major directions on how to encode these information effectively for listwise learning: (1) Metric-based optimization, which attempts to optimize smooth surrogate functions of ranking metrics, such as ApproxNDCG [27] and SoftRank [32]. (2) Probability-based losses, which assume the ranking process follows certain probabilistic patterns. Specifically, ListMLE [36] views the ranking problem as a sequential learning process, maximizing the probability of the observed permutation directly. ListNet [4] defines a cross entropy loss between two probability distributions of permutations that are transformed from the model predictions and the ground truth. However, the existing listwise ranking losses mainly operate on the given list without further inspection.

If we treat the candidate list in the listwise learning as a special "physical system" where different document candidates are regarded as "particles" to compete with each other to be more comparatively relevant to the given query, many fascinating and profound concepts in physics can provide us refreshing alternatives to handle the ranking problem. Coarse graining is an essential paradigm in biophysics. It aims at constructing simplified representations of complex systems so that the major chemical or physical characteristics are preserved and less consequential interactions can be ignored. As a result, it enables us to conduct affordable and reliable simulations to obtain a better understanding on system dynamics. Martin Karplus, Michael Levitt and Arieh Warshel, Nobel laureates in chemistry, introduced the concept of simplification of biomolecular complexes for longer simulations at biological relevant time scales [18, 19, 35]. Their work inspires numerous coarse-grained models for biomolecular modelling [11, 16, 31]. Moreover, the coarse-graining procedure in biophysics modelling can embrace a wide range of length scales, from a few atoms to proteins, nucleic acids, lipid membranes, carbohydrates and water, based on research interests. The same logic can be applied to listwise learning in information retrieval. Not all the documents in the candidate list are pivotal to the learning process of neural retrieval models. In this paper, we propose ExpertRank, a novel expert-based listwise ranking loss, which is built on the concept of multi-level coarse graining and applies the MoE technique to combine semantic matching information at various levels effectively.

## 3 EXPERT-BASED LISTWISE RANKING LOSS

The goal of information retrieval systems is to recommend the most relevant documents for a given query. While semantic relatedness between the query and each individual document is the foundation of relevance estimation, an explicit inter-document comparison is necessary and effective to refine the exact ranking order among candidates which may share similar content signals. In general, listwise learning often outperforms pairwise/pointwise learning due to its innate ability to differentiate various candidates simultaneously on a large scale. This section describes ExpertRank, a new listwise ranking loss, based on the concept of the multi-level coarse graining by applying the technique of mixture of experts (MoE) to the domain of learning to rank. The overall architecture of ExpertRank is displayed in Figure 1.

### 3.1 Coarse Graining

Given a query $q$ and a list of documents, containing $N$ relevant documents $d_1^+, ..., d_N^+$ and $M$ non-relevant documents $d_1^-, ..., d_M^-$, the traditional listwise ranking losses treat the candidate list as a whole and build further analysis upon it. However, some salient signals from certain candidates may be ignored by the other more prominent candidates. Especially for a large number of non-relevant documents $M$, the noise from candidate peers may overshadow some nuanced semantic information due to candidate list quality. Inspired by its success in biophysical modelling, the core idea of coarse graining within ExpertRank is to build a simplified representation of the orginal candidate list by selecting those pronounced candidates at a local level. Specifically, when the list size is relatively small, documents with moderate yet efficacious signals are more likely to stand out. The initial local-level comparison can also filter out those less interesting candidates. The size of local filtering controls the granularity of coarse graining. Coarse graining at different granularity level can yield different useful signals from those refined candidate lists.

How to conduct local filtering effectively across different scales for the given candidate list is crucial to the success of ExpertRank. In most realistic applications, the number of non-relevant documents is far larger than that of relevant documents. Often there may even be only one or two relevant documents available in a given dataset. Hence, coarse graining will only be applied to non-relevant documents, while relevant documents always remain. Moreover, due to its non-differentiability, the direct sorting is not an option to compare a sub-group of documents for neural retrieval models that depend on back-propagation optimization. Therefore, ExpertRank employs the classic pooling technique to conduct local filtering and to build coarse-grained candidate lists. The simplicity and differentiability of the pooling technique enable it to be easily scalable to various levels for coarse graining.

Coarse graining at different levels offers vastly different granularities. If the pooling window is too small, the difference within each sub-group of candidates may be unstable and less significant. Conversely, too few documents can be selected to provide enough useful information if the pooling window is relatively large. Hence, to incorporate both scenarios, ExpertRank identifies two promising ranges of pooling sizes, Low and High, compared to the given candidate list size of listwise learning. Then it selects two different pooling sizes from each range to make the overall learning process more stable.

Specifically, within each pooling window with the low-range pooling sizes, only the document with the maximum predicted ranking score is selected. Since coarse graining is only conducted on the non-relevant documents, candidates with the local maximum scores show more partial relevance than others to the given query. Those local maximum candidates are more important and challenging to ensure the final training performance, since there are more chance to confuse them with the truly relevant documents due to their partial relevance. With the help of coarse graining, we are able to extract and compare those partially relevant candidates with the truly relevant documents directly to aid model learning. On the other hand, within each pooling window with the high-range pooling sizes, both documents with maximum/minimum scores are chosen to provide adequate candidate information. Candidates with the local minimum scores can be more distinguishably non-relevant to inform models of the negative information. As a result, they are useful candidates to subsidize the scarce local maximum documents in this scenario.

### 3.2 Expert Learning

For each coarse-grained candidate list, ExpertRank applies listwise learning separately. There are many possible choices from the traditional listwise learning schemes. However, metric-based listwise ranking losses, such as ApproxNDCG [27] and SoftRank [32], tend to be trapped in the local minima due to its non-convex nature, which may prevent it from being robust and applicable to complex retrieval models. On the other hand, ListMLE [36] regards the ranking problem as a sequential selection process. This assumption yields a rather complicated loss function to optimize, which makes the overall training process delicate and slow. Therefore, ExpertRank chooses ListNet [4] as the backbone of expert learning on account of its simple probability-based loss formulation and efficient convergence rate. ListNet on each coarse-grained candidate list is regarded as an "Expert". Then, the technique of mixture of experts (MoE) is employed to combine different local experts for the final learning signal effectively.

### 3.3 ExpertRank

Like the general mixture of experts (MoE) technique [12, 15, 30], ExpertRank consists of two major components, expert learning and a gating network. The gating network utilises both the matching features from the neural retrieval model and prediction score ranges within each pooling window to generate a probability distribution over different expert outputs. Within each pooling window, the average positive score is $\overline{S^+}$, the max/min non-relevant scores are $S^-_{max}$ and $S^-_{min}$. We then compute three score ranges for gating signals, (1) $\overline{S^+} - S^-_{max}$, (2) $\overline{S^+} - S^-_{min}$, (3) $S^-_{max} - S^-_{min}$, which provides insightful information about prediction variance within each window. Moreover, by feeding matching features associated with the selected scores into gating network, ExpertRank can further aid parameter learning during model training.

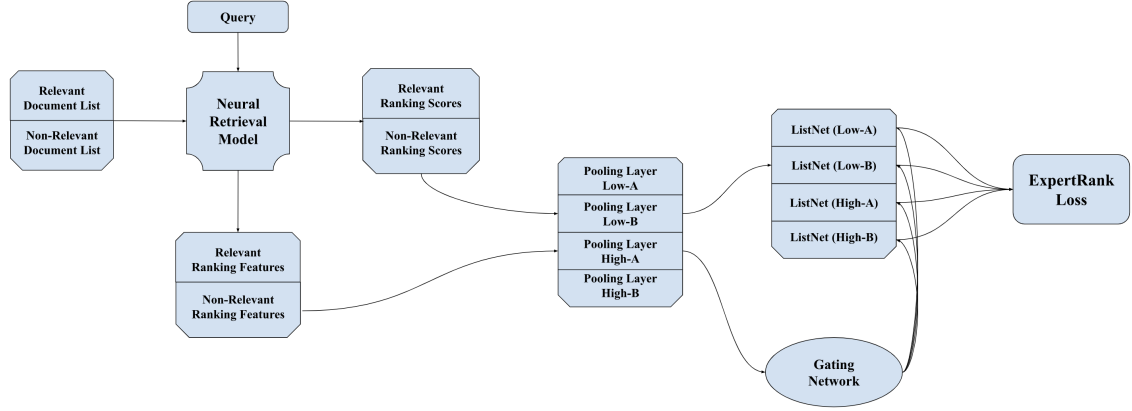

Figure 1: The overall architecture of the Expert-based Listwise Ranking Loss, ExpertRank. Relevant/Non-relevant features are the matching features from the neural retrieval model corresponding to relevant/non-relevant scores. Low-A/B and High-A/B refer to two different pooling layers with small and large pooling sizes, respectively.

Mathematically, given a query $q$ and a list of documents, containing $N$ relevant documents $\{d^+_i\}^N_{i=1}$ and $M$ non-relevant documents $\{d^-_i\}^M_{i=1}$, the predicted ranking scores of relevant documents are $\{S^+_i\}^N_{i=1}$ and the corresponding matching features are $\{F^+_i\}^N_{i=1}$. For the experts with a low-range pooling size $p_L$, a max-pooling layer is applied to select more prominent non-relevant candidates with scores $\{_{max}S^-_i\}^{M_L}_{i=1}$ and matching features $\{_{max}F^-_i\}^{M_L}_{i=1}$. The associated experts can be written as

$$E_L = ListNet\left(\{S^+_i\}^N_{i=1},\ \{_{max}S^-_i\}^{M_L}_{i=1}\right) \quad (1)$$

The gating signals associated with score ranges and matching features can be written as $g^s_L$ and $g^f_L$, respectively, where $g^s_L$ and $g^f_L$ are two hidden layers to encode the corresponding information:

$$\begin{aligned} g^s_L &= g^s_L\left(\{S^+_i\}^N_{i=1},\ \{S^-_i\}^M_{i=1}\right) \\ g^f_L &= g^f_L\left(\{F^+_i\}^N_{i=1},\ \{_{max}F^-_i\}^{M_L}_{i=1}\right) \end{aligned} \quad (2)$$

Similarly, for the experts with a high-range pooling size $p_H$, a pair of max/min-pooling layers is applied to select non-relevant candidates with scores $\{_{max}S^-_i\}^{M_H}_{i=1}, \{_{min}S^-_i\}^{M_H}_{i=1}$ and matching features $\{_{max}F^-_i\}^{M_H}_{i=1}, \{_{min}F^-_i\}^{M_H}_{i=1}$. The associated experts can be written as

$$E_H = ListNet\left(\{S^+_i\}^N_{i=1},\ \{_{max}S^-_i\}^{M_H}_{i=1},\ \{_{min}S^-_i\}^{M_H}_{i=1}\right) \quad (3)$$

The gating signals associated with score ranges and matching features can be written as $g^s_H$ and $g^f_H$, respectively:

$$\begin{aligned} g^s_H &= g^s_H\left(\{S^+_i\}^N_{i=1},\ \{S^-_i\}^M_{i=1}\right) \\ g^f_H &= g^f_H\left(\{F^+_i\}^N_{i=1},\ \{_{max}F^-_i\}^{M_H}_{i=1},\ \{_{min}F^-_i\}^{M_H}_{i=1}\right) \end{aligned} \quad (4)$$

The gating network then can be written as

$$G = Softmax\left(\{g^s_L\}^2_{L=1}, \{g^f_L\}^2_{L=1}, \{g^s_H\}^2_{H=1}, \{g^f_H\}^2_{H=1}\right) \quad (5)$$

Finally, ExpertRank loss can be presented as

$$L_{ExpertRank} = \sum_{i=1}^{4} G_i E_i \quad (6)$$

It is worth noting that the MoE layer makes no changes on the structure of neural retrieval models, even though it aims to aid model training. Once listwise learning is complete, the learnable parameters associated with ExpertRank can be removed during model evaluation. The neural retrieval models can be used in the same way as before without any modifications, which makes ExpertRank highly generalizable and applicable to various systems.

## 4 NEURAL RETRIEVAL MODEL SELECTION

In general, the performance of deep learning models heavily relies on model complexity, data quality and effective model training. Given the same dataset available for model training, model complexity is a strong indicator of model potential. The major purpose of learning-to-rank losses is to exploit their full potential via feeding salient signals to neural networks. Hence, we select different neural retrieval models across different complexity levels to demonstrate the robustness, effectiveness and generalizability of ExpertRank.

**KNRM** [5] constructs a similarity matrix computed by the word embeddings of the given query and document terms. The model then extracts multi-level soft-match features via Gaussian kernel pooling technique. The multi-level soft-match features enter a hidden layer to compute the final ranking score. ExpertRank uses those critical multi-level soft-match features for the expert gating network.

**ConvKNRM** [38] is an enhanced extension of KNRM by computing n-gram embeddings via a Convolutional Neural Network (CNN) laye. Similarly, Gaussian kernel pooling technique is applied to generate the multi-level soft-match features, which is used for ExpertRank. Its final ranking score is the output of the final hidden layer fed by those multi-level soft-match features.

**MatchPyramid** [23] treats the constructed similarity matrix from query and document word embeddings as a special "image". The model then applies several stacked CNN layers to extract distinct signals from the similarity matrix image, which is a popular choice for many neural-based computer vision architectures. The CNN-based features then are used to compute the final ranking score via several stacked hidden layers. ExpertRank relies on those CNN-based features for expert gating network.

**BERT** [8] model has been very a popular choice to encode text information ever since it's proposed. BERT is the first deeply bidirectional, unsupervised language representation, pre-trained using only a plain text corpus. Due to its powerful text encoding capability and promising performance in many natural language processing tasks, various BERT-based retrieval models have been proposed [1, 26]. Qian *et al.* [26] conduct extensive experiments to apply BERT to document ranking using different embedding outputs and different fine-tuned layers. Based on their results, we choose the sentence-level BERT embedding of query-document concatenation as the matching signals, which is also used for expert gating network within ExpertRank. Several hidden layers are then applied to compute the final ranking score.

**ALBERT** model [17] is proposed to to optimize performance and reduce model parameters of BERT model. By utilising factorization of the embedding parametrization and cross layer parameter sharing, an ALBERT-base model that has only $12M$ parameters, an 89% parameter reduction compared to the BERT-base model, yet still achieves respectable performance across the benchmarks considered. We then replace BERT-based embedding with ALBERT-based embedding as the matching signals for document ranking.

### 4.1 Listwise Learning for BERT/ALBERT-IR

Listwise learning often provides more reliable and competitive performance than pointwise learning and pairwise learning, since its foundation of inter-candidate comparison is more consistent with the comparison-based nature of document ranking problem. However, one major drawback of listwise learning is the relatively high computing memory requirement, especially for neural retrieval models. Given a list size $L$ during listwise model training, $L$ query-document pairs are fed into the system all together as a single list, which requires significantly more computing memory. To address this issue, instead of using trainable BERT/ALBERT models, fixed pre-trained BERT/ALBERT embeddings are pre-computed for each query-document pair. The simple pointwise classification loss is applied to fine-tune BERT/ALBERT-based retrieval model on the given training and validation dataset to construct the fixed fine-tuned BERT/ALBERT embeddings. Then only several hidden layers are trained to compute ranking scores during listwise learning. To ensure credibility of the further listwise re-training, both classification loss and listwise ranking losses utilize the same training and validation dataset. Test dataset is only used to evaluate model performance, not for fine-tuning.

### 4.2 Model Implementation Setting

All our neural retrieval models are implemented in tensorflow[1]. For KNRM, ConvKNRM and MatchPyramid, GloVe word embeddings with 300 dimensions [25], trained on the Common Crawl dataset[2], are used for vocabulary initialization. For BERT and ALBERT, the corresponding tokenizers from *bert-for-tf2* package[3] are used to preprocess the raw text data, respectively. We use an Adam optimizer to train all models. The learning rate is set to 1e−4 for all the listwise trainings, and to 1e−6 for BERT and ALBERT fine tuning.

Regarding model architecture parameters, for both KNRM and ConvKNRM, we apply 11 Gaussian kernels with one harvesting exact matches (mean value $\mu_0 = 1.0$ and standard deviation $\sigma_0 = 10^{-3}$) and ten capturing soft matches (mean values spaced evenly in the range of $[-1, 1]$, from -0.9 to 0.9, and standard deviation $\sigma_1 = 0.1$). The number of CNN filters in ConvKNRM is set to 128. For MatchPyramid, we choose 3 convolution layers, each with kernel size $3 \times 3$ and 16 convolutional channels. We choose the pre-trained uncased BERT-base model and ALBERT-base model, both with 12 hidden layers and 12 attention heads, for BERT/ALBERT-based retrieval models.

## 5 EXPERIMENT DESIGN

Various empirical experiments are conducted to further validate the robustness, effectiveness and generalizability of ExpertRank. All the selected neural retrieval models are trained via ExpertRank, along with other existing listwise ranking losses, such as ListNet [4], ListMLE [36] and ApproxNDCG [27].

### 5.1 Performance on MS MARCO

Deep learning models are known to be data-hungry. Especially, models with high complexity tend to require more data for training. Effective and robust learning mechanism aids models in reaching their full potential. The MS MARCO (Microsoft Machine Reading Comprehension) passage ranking dataset [24] is a well-known large-scale public dataset, containing nearly 550,000 unique queries from the Bing search engine and more than 8.8 million unique passages. This collection consists of the `top1000` training dataset and the `top1000` development dataset, which are pre-selected using BM25 [29] for the given query. In our experiments, we randomly choose 20% of the development set (1,396 queries) for model validation, and the remaining 80% (5,583 queries) as test data. Furthermore, MS MARCO is a raw text dataset. The maximum word length of queries and documents is set to 20 and 180, which covers more than 99% cases. Due to the corresponding high memory requirement for listwise learning, we randomly select 50 non-relevant documents for each query from the `top1000` training dataset, while relevant documents are always selected, since most queries have only single positive passage. For the validation and test datasets, `top1000` passages for each query are used for model evaluation. We evaluate and compare different listwise ranking losses in terms of the Mean Reciprocal Rank(MRR), the Normalized Discounted Cumulative Gain (nDCG), and the full Mean Average Precision (MAP). The test set performance comparison is shown in Table 1.

---
[1] https://www.tensorflow.org/
[2] https://commoncrawl.org/
[3] https://pypi.org/project/bert-for-tf2/0.1.1/

Table 1: Test Performance on MS MARCO dataset. *P* refers to the best selected pooling window size combination in ExpertRank. BERT-raw and ALBERT-raw refer to the retrieval models using the fixed pre-trained BERT/ALBERT embeddings, respectively. BERT-trained and ALBERT-trained refer to the retrieval models using the fixed fine-tuned embeddings, respectively. BERT-trained_CE and ALBERT-trained_CE indicate the initial fine-tuning performance using cross entropy loss before the further listwise training. ¶ indicates statistically significant improvements over all the other losses. § indicates full statistically significant improvement except ApproxNDCG.

| Model | MRR@3 | MRR@10 | MRR | nDCG@3 | nDCG@10 | nDCG | MAP@3 | MAP@10 | MAP | P |
|---|---|---|---|---|---|---|---|---|---|---|
| KNRM_ListMLE | 0.18559 | 0.22015 | 0.23331 | 0.20312 | 0.27432 | 0.36866 | 0.18191 | 0.21651 | 0.23008 | |
| KNRM_ListNet | 0.19150 | 0.22438 | 0.23821 | 0.20943 | 0.27758 | 0.37343 | 0.18751 | 0.22081 | 0.23493 | |
| KNRM_nDCG | 0.18538 | 0.21723 | 0.23056 | 0.20208 | 0.26852 | 0.36517 | 0.18113 | 0.21353 | 0.22715 | |
| KNRM_ExpertRank | **0.20106¶** | **0.23464¶** | **0.24809¶** | **0.21911¶** | **0.28954¶** | **0.38268¶** | **0.19640¶** | **0.23048¶** | **0.24430¶** | [5, 7, 10, 25] |
| ConvKNRM_ListMLE | 0.22082 | 0.25414 | 0.26712 | 0.23871 | 0.30792 | 0.39845 | 0.21601 | 0.24945 | 0.26296 | |
| ConvKNRM_ListNet | 0.22709 | 0.26115 | 0.27485 | 0.24467 | 0.31534 | 0.40650 | 0.22263 | 0.25677 | 0.27095 | |
| ConvKNRM_nDCG | 0.16998 | 0.20383 | 0.21911 | 0.18518 | 0.25538 | 0.35855 | 0.16613 | 0.20000 | 0.21567 | |
| ConvKNRM_ExpertRank | **0.24906¶** | **0.28576¶** | **0.29913¶** | **0.26755¶** | **0.34345¶** | **0.42892¶** | **0.24344¶** | **0.28053¶** | **0.29447¶** | [4, 7, 10, 25] |
| MatchPyramid_ListMLE | 0.19052 | 0.21803 | 0.23198 | 0.20538 | 0.26262 | 0.36345 | 0.18649 | 0.21406 | 0.22824 | |
| MatchPyramid_ListNet | 0.22503 | 0.25825 | 0.27196 | 0.24251 | 0.31173 | 0.40343 | 0.22001 | 0.25358 | 0.26776 | |
| MatchPyramid_nDCG | 0.22031 | 0.25402 | 0.26832 | 0.23847 | 0.30861 | 0.40162 | 0.21556 | 0.24933 | 0.26419 | |
| MatchPyramid_ExpertRank | **0.23712¶** | **0.27224¶** | **0.28568¶** | **0.25548¶** | **0.32794¶** | **0.41634¶** | **0.23192¶** | **0.26714¶** | **0.28110¶** | [5, 7, 10, 25] |
| BERT-raw_ListMLE | 0.10466 | 0.13273 | 0.14950 | 0.11556 | 0.17437 | 0.29426 | 0.10184 | 0.12984 | 0.14701 | |
| BERT-raw_ListNet | 0.14016 | 0.17036 | 0.18609 | 0.15420 | 0.21770 | 0.32784 | 0.13617 | 0.16647 | 0.18264 | |
| BERT-raw_nDCG | 0.14132 | 0.17182 | 0.18757 | 0.15450 | 0.21834 | 0.32876 | 0.13713 | 0.16763 | 0.18387 | |
| BERT-raw_ExpertRank | **0.15526¶** | **0.18580¶** | **0.20161¶** | **0.16953¶** | **0.23355¶** | **0.34166¶** | **0.15076¶** | **0.18156¶** | **0.19782¶** | [3, 4, 10, 17] |
| BERT-trained_CE | 0.33023 | 0.37060 | 0.38224 | 0.35517 | 0.43962 | 0.50610 | 0.32365 | 0.36501 | 0.37748 | |
| BERT-trained_ListMLE | 0.32936 | 0.37074 | 0.38215 | 0.35422 | 0.44053 | 0.50618 | 0.32314 | 0.36536 | 0.37760 | |
| BERT-trained_ListNet | 0.33154 | 0.37249 | 0.38406 | 0.35656 | 0.44189 | 0.50774 | 0.32521 | 0.36693 | 0.37932 | |
| BERT-trained_nDCG | 0.33426 | 0.37335 | 0.38511 | 0.35943 | 0.44130 | 0.50814 | 0.32744 | 0.36740 | 0.37998 | |
| BERT-trained_ExpertRank | **0.33757¶** | **0.37683¶** | **0.38855¶** | **0.36289¶** | **0.44499¶** | **0.51120¶** | **0.33075¶** | **0.37104¶** | **0.38359¶** | [5, 7, 17, 25] |
| ALBERT-raw_ListMLE | 0.17255 | 0.20398 | 0.22002 | 0.18795 | 0.25284 | 0.35893 | 0.16870 | 0.20001 | 0.21658 | |
| ALBERT-raw_ListNet | 0.20228 | 0.23659 | 0.25151 | 0.21905 | 0.29083 | 0.38784 | 0.19742 | 0.23173 | 0.24734 | |
| ALBERT-raw_nDCG | 0.19733 | 0.23134 | 0.24680 | 0.21440 | 0.28450 | 0.38378 | 0.19307 | 0.22695 | 0.24289 | |
| ALBERT-raw_ExpertRank | **0.21670¶** | **0.25175¶** | **0.26605¶** | **0.23438¶** | **0.30751¶** | **0.40068¶** | **0.21191¶** | **0.24702¶** | **0.26195¶** | [3, 4, 17, 25] |
| ALBERT-trained_CE | 0.34002 | 0.38192 | 0.39350 | 0.36402 | 0.45051 | 0.51536 | 0.33206 | 0.37498 | 0.38745 | |
| ALBERT-trained_ListMLE | 0.34826 | 0.38951 | 0.40094 | 0.37295 | 0.45869 | 0.52195 | 0.34065 | 0.38289 | 0.39521 | |
| ALBERT-trained_ListNet | 0.35092 | 0.39310 | 0.40444 | 0.37590 | 0.46296 | 0.52503 | 0.34350 | 0.38662 | 0.39869 | |
| ALBERT-trained_nDCG | 0.35566 | 0.39757 | 0.40876 | 0.37990 | 0.46675 | 0.52841 | 0.34814 | 0.39100 | 0.40304 | |
| ALBERT-trained_ExpertRank | **0.36005¶** | **0.40180¶** | **0.41280¶** | **0.38420¶** | **0.47059¶** | **0.53153¶** | **0.35223§** | **0.39510¶** | **0.40693¶** | [3, 5, 17, 25] |

ExpertRank outperforms all existing listwise ranking losses consistently and significantly across different neural retrieval models and different ranking metrics, MRR, nDCG and MAP, with different thresholds. Statistical significance tests indicated in Table 1 are done using a two sided paired $t$-test ($p < 0.05$). As the backbone of ExpertRank, ListNet produces competitive and robust results, compared to ListMLE and ApproxNDCG. Specifically, ListMLE fails to yield reliable performance on MatchPyramid and BERT/ALBERT-raw models. On the other hand, although ApproxNDCG shows reasonable results on BERT/ALBERT-related models, its performance on ConvKNRM trails behind other ranking losses by a large margin. Due to its non-convex nature, ApproxNDCG can easily be trapped in the local minima during back propagation optimization, resulting in unstable performance. It is worth noting that fixed embeddings are used in BERT/ALBERT-related retrieval models. During the further listwise learning, the trainable network structures within those models are purposefully simple, which may help Approx-NDCG to avoid being stuck in the sub-optimal positions. However, it may fail to generate reliable results on more complex models, such as ConvKNRM and BERT/ALBERT-based models with trainable embeddings. Hence, the selection of ListNet within ExpertRank is validatable and plausible. Moreover, compared to ListNet, ExpertRank utilizes multi-level coarse-grained information to yield much more competitive performance consistently in various settings.

Regarding the results of BERT/ALBERT-related models, the performance gains from ExpertRank are much more substantial on the models with the raw pre-trained embeddings. Since the raw BERT/ALBERT embeddings are not summarised and learnt from MS MARCO dataset, the performance of BERT/ALBERT-raw models heavily rely on the simple hidden layers which are used to compute the final ranking scores. Results in Table 1 demonstrate the strong capability of ExpertRank to exploit every usable signal even from the simple structure. Furthermore, the fine-tuning of BERT/ALBERT on MS MARCO dataset helps to extract and summarise all the salient information into the fine-tuned embeddings. Hence, the hidden layers in BERT/ALBERT-trained models contribute less for the final ranking decision, which limits the further listwise learning. Nonetheless, ExpertRank still performs better than the initial fine-tuned BERT/ALBERT-trained_CE models. Thus, we believe the performance improvement from ExpertRank on BERT/ALBERT-related retrieval models can be much more pronounced if applied directly with the fine-tuning of trainable embeddings.

## 5.2 Limited Training Data

MS MARCO is a comparably large collection with sufficient training information. One major issue associated with this dataset is its shallow relevance judgment, where the majority of the given queries only have a single relevant document. In this section, we test ExpertRank and other listwise ranking losses in a different scenario. To further validate the effectiveness and robustness of ExpertRank, we conduct our experiments on a proprietary news collection, which contains about 730$k$ English news articles covering various topics. 35 request queries are constructed based on this collection, among which 30 queries are available for model training and the remaining 5 requests are used for evaluation. On average, each request query has around 50 relevant documents identified by a large pool of in-house expert annotators. To aid model training and evaluation, BM25 [29] is applied to pre-select the top-5000 documents for each query. Since the available training information is rather limited, we train each neural retrieval model for a fixed 15 epochs and save trained models every 3 epochs. Then evaluation is carried out on each saved model to select the best one. We use the same model settings as in the previous MS MARCO experiments. For simplicity, we reuse the pooling size combinations of ExpertRank associated with each model determined on MS MARCO. We evaluate and compare different listwise ranking losses in terms of Precision@10, Recall@10, and nDCG@10. Test performance is shown in Table 2.

Compared to other listwise ranking losses, ExpertRank gives more competitive and reliable performance consistently across different neural retrieval models. Although ListMLE yields reasonable results for KNRM and ConvKNRM, it fails to train MatchPyramid effectively with the limited training data. Its performance on BERT/ALBERT-based models also pales in comparison with other ranking losses. On the other hand, ApproxNDCG cannot provide reliable results for KNRM and ConvKNRM. Its performance leads on BERT/ALBERT-based models are not consistent with its results on MS MARCO, which indicates its strong dependence on the amount and quality of training. ListNet demonstrates resilient performance with respect to ListMLE and ApproxNDCG. By combining pooling-based Mixture of Experts (MoE) and ListNet, ExpertRank yields much more competitive results even with limited training information. Its performance on this news dataset further demonstrates the potential usage of ExpertRank in various realistic domains. It is worth mentioning that for convenience, we used the same pooling size combinations that we earlier determined MS MARCO. We believe that even more prominent gains can be achieved by dedicated fine-tuning on this collection.

Table 2: Test Performance on the news collection.

| Model | Precision@10 | Recall@10 | nDCG@10 |
|---|---|---|---|
| KNRM_ListMLE | 0.26 | 0.01761 | 0.25728 |
| KNRM_ListNet | 0.18 | 0.01098 | 0.20760 |
| KNRM_nDCG | 0.18 | 0.01143 | 0.16119 |
| KNRM_ExpertRank | **0.28** | **0.01864** | **0.28239** |
| ConvKNRM_ListMLE | 0.90 | 0.06089 | 0.90519 |
| ConvKNRM_ListNet | 0.90 | 0.06151 | **0.92220** |
| ConvKNRM_nDCG | 0.84 | 0.05605 | 0.87346 |
| ConvKNRM_ExpertRank | **0.92** | **0.06344** | 0.89295 |
| MatchPyramid_ListMLE | 0.42 | 0.02780 | 0.51918 |
| MatchPyramid_ListNet | 0.84 | 0.05728 | 0.88626 |
| MatchPyramid_nDCG | 0.84 | 0.05705 | 0.83262 |
| MatchPyramid_ExpertRank | **0.90** | **0.06232** | **0.91341** |
| BERT-raw_ListMLE | 0.54 | 0.04076 | 0.51339 |
| BERT-raw_ListNet | 0.56 | 0.04175 | 0.58234 |
| BERT-raw_nDCG | **0.60** | 0.04355 | 0.62708 |
| BERT-raw_ExpertRank | **0.60** | **0.04410** | **0.63090** |
| BERT-trained_ListMLE | 0.72 | 0.04881 | 0.72634 |
| BERT-trained_ListNet | 0.86 | 0.05973 | 0.85983 |
| BERT-trained_nDCG | 0.78 | 0.05354 | 0.77590 |
| BERT-trained_ExpertRank | **0.90** | **0.06160** | **0.89179** |
| ALBERT-raw_ListMLE | 0.78 | 0.05447 | 0.81848 |
| ALBERT-raw_ListNet | 0.80 | 0.05489 | 0.79763 |
| ALBERT-raw_nDCG | 0.76 | 0.05272 | 0.79373 |
| ALBERT-raw_ExpertRank | **0.84** | **0.05818** | **0.85929** |
| ALBERT-trained_ListMLE | 0.86 | 0.05837 | 0.89266 |
| ALBERT-trained_ListNet | 0.90 | 0.06147 | 0.92050 |
| ALBERT-trained_nDCG | 0.82 | 0.05540 | 0.81492 |
| ALBERT-trained_ExpertRank | **0.92** | **0.06271** | **0.92926** |

## 6 POOLING SIZE ANALYSIS

Choosing pooling size combinations properly is crucial to the success of ExpertRank. Different pooling sizes conduct local comparison at different levels for coarse graining. Those selected candidates then are utilised to feed each ListNet-based "expert" learner. Given the same data quality, the final ranking performance of neural retrieval models depends on model potential and effective training. Models with different neural network complexities have different innate abilities to distinguish document candidates at different granularity levels. Hence, the optimal pooling size combination may vary from case to case.

Here, we choose the number of non-relevant documents to be 50. Then ExpertRank identifies typical basis pooling sizes, {2, 3, 4, 5, 7} for the low-range and {10, 17, 25} for the high-range. Two pooling sizes are selected from each range, respectively. In Figure 2, we demonstrate how the performance of each retrieval model changes with respect to pooling size combination on MS MARCO collection. We select 15 typical pooling size combinations across different models. Other combinations yield similar performance trends. The dashed lines show the performance of other listwise baselines we tested. Also, since the majority of queries in MS MARCO only have a single relevant document, metrics of MRR, nDCG, MAP become approximately proportional to each other and yield similar trends. Thus, only MRR@10 is reported in Figure 2.

Compared to the existing listwise baselines, ExpertRank with different pooling size combinations yields better performance consistently across various retrieval models. Especially for KNRM, ConvKNRM, MatchPyramid and BERT_raw, the gap between ExpertRank and ListMLE/ListNet is consistent and significant. On the other hand, the majority of pooling size combinations for BERT_trained and ALBERT_raw show reliable performance and consistent effectiveness. ALBERT_trained model tends to prefer higher end pooling sizes. As mentioned in Section 4, BERT/ALBERT_trained models apply fixed fine-tuned embeddings for further listwise re-training. The structure of models such as KNRM, ConvKNRM and MatchPyramid is considerably simpler leading to their somewhat smaller learning potential. We believe that even more significant performance gains can be achieved in the future if the trainable BERT/ALBERT embeddings were employed directly in the listwise training.

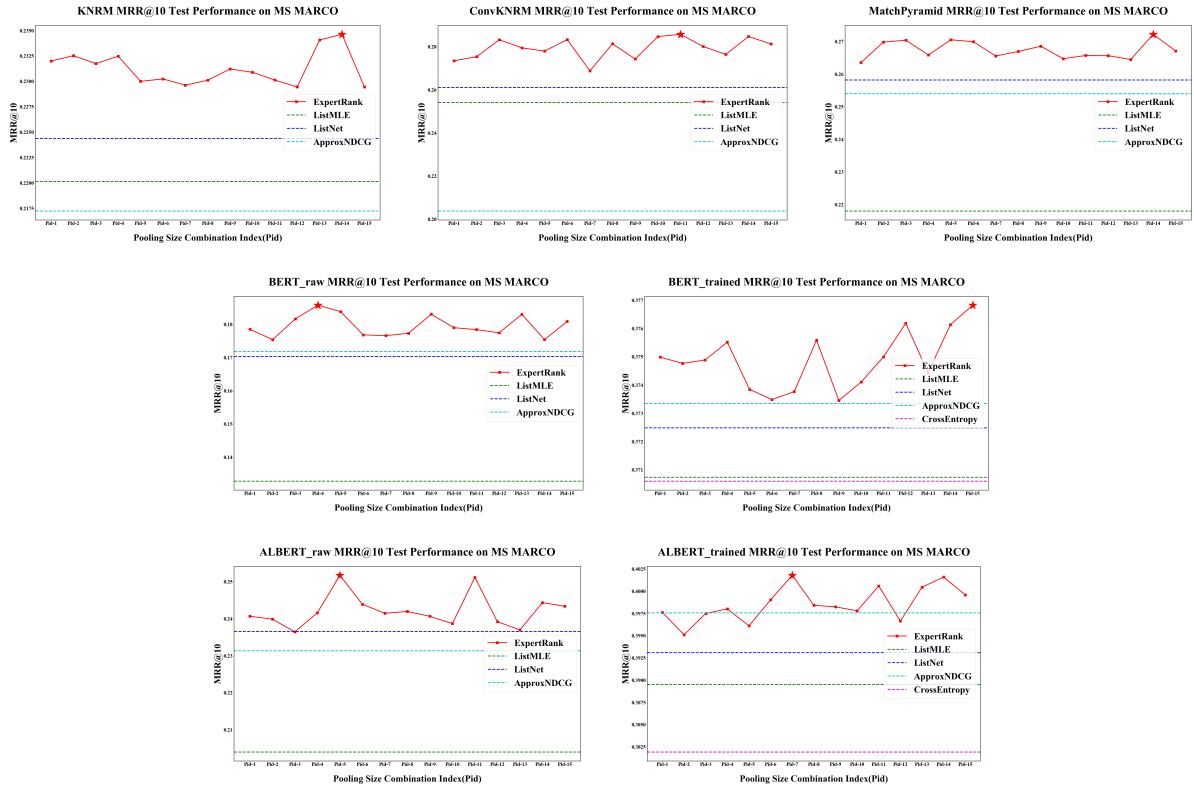

Figure 2: ExpertRank MRR@10 test performance on MS MARCO with different pooling size combinations. ★ indicates the best performance. The horizontal lines refer to ListMLE, ListNet, ApproxNDCG and Cross Entropy(for the initial BERT/ALBERT fine-tuning), respectively. The 15 Pooling Size Combination Indices refer to [2, 3, 10, 17], [2, 3, 10, 25], [2, 7, 10, 17], [3, 4, 10, 17], [3, 4, 17, 25], [3, 5, 10, 17], [3, 5, 17, 25], [3, 7, 10, 17], [4, 5, 10, 17], [4, 7, 10, 17], [4, 7, 10, 25], [4, 7, 17, 25], [5, 7, 10, 17], [5, 7, 10, 25], [5, 7, 17, 25], respectively.

## 7 CONCLUSION

The mixture of experts (MoE) technique is a powerful mechanism that can combine different "sub-learners" efficiently to produce better performance. In this paper, we present ExpertRank, an expert-based listwise ranking loss for retrieval. Instead of treating the candidate list as a whole like other listwise ranking losses, ExpertRank regards different documents as "particles" within a special "physical system". Then it utilizes the classic pooling technique to conduct multi-level coarse graining so that candidates with moderate relevance estimates can still stand out. The local-level comparison provided by the pooling layers can filter out certain less interesting documents and force the listwise learning to focus more on the proper candidates. By employing the MoE technique, ExpertRank combines listwise learning on various scales effectively. As a result, ExpertRank is highly robust and reliably generalizable to arbitrary neural retrieval models. In the experiments on MS MARCO and a proprietary news collection, we demonstrate that ExpertRank outperforms other existing listwise ranking losses consistently across different models and provides a strong learning alternative. There are several promising directions for future work. We are eager to investigate whether hierarchical MoE techniques can further improve the ranking performance. Moreover, ExpertRank relies on a simple pooling technique and the traditional ListNet scheme to construct each "expert". More well-tuned sub-learner structures may further boost the overall performance.

## ACKNOWLEDGMENTS

Acknowledgements blinded for anonymous review.

# REFERENCES

[1] Zeynep Akkalyoncu Yilmaz, Shengjin Wang, Wei Yang, Haotian Zhang, and Jimmy Lin. 2019. Applying BERT to Document Retrieval with Birch. In *Proceedings of the 2019 Conference on Empirical Methods in Natural Language Processing and the 9th International Joint Conference on Natural Language Processing (EMNLP-IJCNLP): System Demonstrations*. Association for Computational Linguistics, Hong Kong, China, 19–24. https://doi.org/10.18653/v1/D19-3004

[2] Christopher Burges, Robert Ragno, and Quoc Le. 2006. Learning to Rank with Nonsmooth Cost Functions. 193–200.

[3] Chris Burges, Tal Shaked, Erin Renshaw, Ari Lazier, Matt Deeds, Nicole Hamilton, and Greg Hullender. 2005. Learning to Rank Using Gradient Descent. In *Proceedings of the 22nd International Conference on Machine Learning* (Bonn, Germany) *(ICML '05)*. Association for Computing Machinery, New York, NY, USA, 89–96. https://doi.org/10.1145/1102351.1102363

[4] Zhe Cao, Tao Qin, Tie-Yan Liu, Ming-Feng Tsai, and Hang Li. 2007. Learning to Rank: From Pairwise Approach to Listwise Approach. In *Proceedings of the 24th International Conference on Machine Learning* (Corvalis, Oregon, USA) *(ICML '07)*. Association for Computing Machinery, New York, NY, USA, 129 – 136. https://doi.org/10.1145/1273496.1273513

[5] Jamie Callan Zhiyuan Liu Chenyan Xiong, Zhuyun Dai and Russell Power. 2017. End-to-end neural ad-hoc ranking with kernel pooling. In *Proceedings of the 40th International ACM SIGIR conference on research and development in information retrieval (SIGIR '17)*.

[6] Daniel Cohen, Scott M. Jordan, and W. Bruce Croft. 2019. Learning a Better Negative Sampling Policy with Deep Neural Networks for Search. In *Proceedings of the 2019 ACM SIGIR International Conference on Theory of Information Retrieval* (Santa Clara, CA, USA) *(ICTIR '19)*. Association for Computing Machinery, New York, NY, USA, 19–26. https://doi.org/10.1145/3341981.3344220

[7] Ronan Collobert, Samy Bengio, and Yoshua Bengio. 2002. A Parallel Mixture of SVMs for Very Large Scale Problems. *Neural Comput.* 14, 5 (May 2002), 1105–1114. https://doi.org/10.1162/089976602753633402

[8] Jacob Devlin, Ming-Wei Chang, Kenton Lee, and Kristina Toutanova. 2019. BERT: Pre-training of Deep Bidirectional Transformers for Language Understanding. In *Proceedings of the 2019 Conference of the North American Chapter of the Association for Computational Linguistics: Human Language Technologies, Volume 1 (Long and Short Papers)*. Association for Computational Linguistics, Minneapolis, Minnesota, 4171–4186. https://doi.org/10.18653/v1/N19-1423

[9] Joshua Goodman. 2001. Classes for Fast Maximum Entropy Training. *Acoustics, Speech, and Signal Processing, 1988. ICASSP-88., 1988 International Conference on* 1, 561 – 564 vol.1. https://doi.org/10.1109/ICASSP.2001.940893

[10] Kai Hui, Andrew Yates, Klaus Berberich, and Gerard de Melo. 2017. PACRR: A Position-Aware Neural IR Model for Relevance Matching. In *EMNLP*.

[11] H. Ingólfsson, Cesar A. López, J. Uusitalo, D. H. de Jong, S. M. Gopal, X. Periole, and S. Marrink. 2014. The power of coarse graining in biomolecular simulations. *Wiley Interdisciplinary Reviews. Computational Molecular Science* 4 (2014), 225 – 248.

[12] R. A. Jacobs, M. I. Jordan, S. J. Nowlan, and G. E. Hinton. 1991. Adaptive Mixtures of Local Experts. *Neural Computation* 3, 1 (1991), 79–87. https://doi.org/10.1162/neco.1991.3.1.79

[13] Qingyao Ai and W. Bruce Croft Jiafeng Guo, Yixing Fan. 2016. A Deep Relevance Matching Model for Ad-hoc Retrieval. In *Proceedings of the 25th ACM International on Conference on Information and Knowledge Management (CIKM '16)*.

[14] Thorsten Joachims. 2002. Optimizing Search Engines Using Clickthrough Data. In *Proceedings of the Eighth ACM SIGKDD International Conference on Knowledge Discovery and Data Mining* (Edmonton, Alberta, Canada) *(KDD '02)*. Association for Computing Machinery, New York, NY, USA, 133–142. https://doi.org/10.1145/775047.775067

[15] Michael I. Jordan and Robert A. Jacobs. 1994. Hierarchical Mixtures of Experts and the EM Algorithm. *Neural Comput.* 6, 2 (March 1994), 181–214. https://doi.org/10.1162/neco.1994.6.2.181

[16] Sebastian Kmiecik, Dominik Gront, Michal Kolinski, Lukasz Wieteska, Aleksandra Badaczewska-Dawid, and Andrzej Kolinski. 2016. Coarse-Grained Protein Models and Their Applications. *Chemical Reviews* 116 (06 2016). https://doi.org/10.1021/acs.chemrev.6b00163

[17] Zhenzhong Lan, Mingda Chen, Sebastian Goodman, Kevin Gimpel, Piyush Sharma, and Radu Soricut. 2020. ALBERT: A Lite BERT for Self-supervised Learning of Language Representations. In *International Conference on Learning Representations*. https://openreview.net/forum?id=H1eA7AEtvS

[18] M. Levitt. 2014. Birth and future of multiscale modeling for macromolecular systems (Nobel Lecture). *Angewandte Chemie* 53 38 (2014), 10006–18.

[19] Michael Levitt and Arieh Warshel. 1975. Computer simulation of protein folding. *Nature* 253 (03 1975), 694–8. https://doi.org/10.1038/253694a0

[20] Zhengdong Lu and Hang Li. 2013. A Deep Architecture for Matching Short Texts. In *Proceedings of the 26th International Conference on Neural Information Processing Systems - Volume 1* (Lake Tahoe, Nevada) *(NIPS'13)*. Curran Associates Inc., Red Hook, NY, USA, 1367–1375.

[21] Jiaqi Ma, Zhe Zhao, Xinyang Yi, Jilin Chen, Lichan Hong, and Ed H. Chi. 2018. *Modeling Task Relationships in Multi-Task Learning with Multi-Gate Mixture-of-Experts*. Association for Computing Machinery, New York, NY, USA, 1930–1939. https://doi.org/10.1145/3219819.3220007

[22] Andriy Mnih and Koray Kavukcuoglu. 2013. Learning Word Embeddings Efficiently with Noise-Contrastive Estimation. In *Proceedings of the 26th International Conference on Neural Information Processing Systems - Volume 2* (Lake Tahoe, Nevada) *(NIPS'13)*. Curran Associates Inc., Red Hook, NY, USA, 2265–2273.

[23] Liang Pang, Yanyan Lan, Jiafeng Guo, Jun Xu, Shengxian Wan, and Xueqi Cheng. 2016. Text Matching as Image Recognition. In *Proceedings of the Thirtieth AAAI Conference on Artificial Intelligence* (Phoenix, Arizona) *(AAAI '16)*. AAAI Press, 2793–2799.

[24] Nick Craswell Li Deng Jianfeng Gao Xiaodong Liu Rangan Majumder Andrew McNamara Bhaskar Mitra Tri Nguyen Mir Rosenberg Xia Song Alina Stoica Saurabh Tiwary Payal Bajaj, Daniel Campos and Tong Wang. 2016. MS MARCO: A Human Generated MAchine Reading COmprehension Dataset. In *30th Conference on Neural Information Processing Systems (NIPS '16)*.

[25] Jeffrey Pennington, Richard Socher, and Christopher D. Manning. 2014. GloVe: Global Vectors for Word Representation. In *Empirical Methods in Natural Language Processing (EMNLP)*. 1532–1543. http://www.aclweb.org/anthology/D14-1162

[26] Yifan Qiao, Chenyan Xiong, Zhenghao Liu, and Zhiyuan Liu. 2019. Understanding the Behaviors of BERT in Ranking. *ArXiv* abs/1904.07531 (2019).

[27] Tao Qin, Tie-Yan Liu, and Hang Li. 2010. A general approximation framework for direct optimization of information retrieval measures. *Information Retrieval* 13, 4 (2010), 375 – 397. https://doi.org/10.1007/s10791-009-9124-x

[28] Carl Edward Rasmussen and Zoubin Ghahramani. 2001. Infinite Mixtures of Gaussian Process Experts. In *Proceedings of the 14th International Conference on Neural Information Processing Systems: Natural and Synthetic* (Vancouver, British Columbia, Canada) *(NIPS'01)*. MIT Press, Cambridge, MA, USA, 881–888.

[29] Stephen Robertson and Hugo Zaragoza. 2009. The Probabilistic Relevance Framework: BM25 and Beyond. *Foundations and Trends in Information Retrieval.* 3, 4 (2009), 333–389. https://doi.org/10.1561/1500000019

[30] Noam Shazeer, Azalia Mirhoseini, Krzysztof Maziarz, Andy Davis, Quoc V. Le, Geoffrey E. Hinton, and Jeff Dean. 2017. Outrageously Large Neural Networks: The Sparsely-Gated Mixture-of-Experts Layer.. In *ICLR (Poster)*. OpenReview.net. http://dblp.uni-trier.de/db/conf/iclr/iclr2017.html#ShazeerMMDLHD17

[31] Nidhi Singh and Wenjin Li. 2019. Recent Advances in Coarse-Grained Models for Biomolecules and Their Applications. *International Journal of Molecular Sciences* 20, 15 (2019). https://doi.org/10.3390/ijms20153774

[32] Michael Taylor, John Guiver, Stephen Robertson, and Tom Minka. 2008. SoftRank: Optimizing Non-Smooth Rank Metrics. In *Proceedings of the 2008 International Conference on Web Search and Data Mining* (Palo Alto, California, USA) *(WSDM '08)*. Association for Computing Machinery, New York, NY, USA, 77 – 86. https://doi.org/10.1145/1341531.1341544

[33] Nicolas Usunier, David Buffoni, and Patrick Gallinari. 2009. Ranking with Ordered Weighted Pairwise Classification. In *Proceedings of the 26th Annual International Conference on Machine Learning* (Montreal, Quebec, Canada) *(ICML '09)*. Association for Computing Machinery, New York, NY, USA, 1057–1064. https://doi.org/10.1145/1553374.1553509

[34] Jun Wang, Lantao Yu, Weinan Zhang, Yu Gong, Yinghui Xu, Benyou Wang, Peng Zhang, and Dell Zhang. 2017. IRGAN: A Minimax Game for Unifying Generative and Discriminative Information Retrieval Models. In *Proceedings of the 40th International ACM SIGIR Conference on Research and Development in Information Retrieval* (Shinjuku, Tokyo, Japan) *(SIGIR '17)*. Association for Computing Machinery, New York, NY, USA, 515–524. https://doi.org/10.1145/3077136.3080786

[35] A. Warshel and M. Levitt. 1976. Theoretical studies of enzymic reactions: dielectric, electrostatic and steric stabilization of the carbonium ion in the reaction of lysozyme. *Journal of molecular biology* 103 2 (1976), 227–49.

[36] Fen Xia, Tie-Yan Liu, Jue Wang, Wensheng Zhang, and Hang Li. 2008. Listwise Approach to Learning to Rank: Theory and Algorithm. In *Proceedings of the 25th International Conference on Machine Learning* (Helsinki, Finland) *(ICML '08)*. Association for Computing Machinery, New York, NY, USA, 1192 – 1199. https://doi.org/10.1145/1390156.1390306

[37] Zhe Zhao, Lichan Hong, Li Wei, Jilin Chen, Aniruddh Nath, Shawn Andrews, Aditee Kumthekar, Maheswaran Sathiamoorthy, Xinyang Yi, and Ed Chi. 2019. Recommending What Video to Watch next: A Multitask Ranking System. In *Proceedings of the 13th ACM Conference on Recommender Systems* (Copenhagen, Denmark) *(RecSys '19)*. Association for Computing Machinery, New York, NY, USA, 43–51. https://doi.org/10.1145/3298689.3346997

[38] Jamie Callan Zhuyun Dai, Chenyan Xiong and Zhiyuan Liu. 2018. Convolutional neural networks for soft-matching n-grams in ad-hoc search. In *Proceedings of the eleventh ACM international conference on web search and data mining (WSDM '18)*.